\documentclass[%
 reprint,
 amsmath,amssymb,
 aps,
 pra,
]{revtex4-1}

\usepackage{multirow}
\usepackage{makecell}
\usepackage{graphicx}
\usepackage{dcolumn}
\usepackage{bm}
\usepackage{balance}
\usepackage{latexsym}
\usepackage{wasysym}
\usepackage{color}

\usepackage{enumerate}

\def\crta{\vrule height1.41ex depth-1.27ex width0.34em}
\def\dj{d\kern-0.36em\crta}
\def\Crta{\vrule height1ex depth-0.86ex width0.4em}
\def\Dj{D\kern-0.73em\Crta\kern0.33em}
\newtheorem{theorem}{Theorem}

\allowdisplaybreaks 

\begin{document}

\title{Vector Generation of Quantum Contextual Sets in Even
Dimensional Hilbert Spaces}

\author{Mladen Pavi{\v c}i{\'c}}
\email{mpavicic@irb.hr}
\homepage{http://www.irb.hr/users/mpavicic}
\affiliation{Department of Physics---Nanooptics, Faculty of Math. and Natural Sci.~I, Humboldt University of Berlin, Germany}
\affiliation{Center of Excellence CEMS, Photonics and Quantum Optics
  Unit, Ru\dj er Bo\v skovi\'c Institute, Zagreb, Croatia.}
\author{Norman D. Megill}
\affiliation{Boston Information Group, Lexington, MA 02420, U.S.A.}

\date{May 1, 2019}


\begin{abstract}
  Recently, quantum contextuality has been proved to be
  the source of quantum computation's power. That, together with
  multiple recent contextual experiments, prompts improving the 
  methods of generation of contextual sets and finding their features.
  The most elaborated contextual sets, which offer blueprints for
  contextual experiments and computational gates, are the
  Kochen--Specker (KS) sets. In this paper, we show a method of vector
  generation that supersedes previous methods. It is implemented by
  means of algorithms and programs that generate hypergraphs embodying
  the Kochen-Specker property and that are designed to be carried out
  on supercomputers. We show that vector component generation of KS
  hypergraphs exhausts all possible vectors that can be constructed
  from chosen vector components, in contrast to previous studies
  that used incomplete lists of vectors and therefore missed a
  majority of hypergraphs. Consequently, this unified method is far
  more efficient for generations of KS sets and their implementation
  in quantum computation and quantum communication. Several new KS
  classes and their features have been found and are elaborated on
  in the paper. Greechie diagrams are discussed. A detailed and
  complete blueprint of a particular 21-11 KS set with a complex
  coordinatization is presented in Appendix \ref{app:1}, in contrast
  to the one from the published version of this paper where only a
  few of its states were given.
\end{abstract}

\maketitle

\section{\label{sec:intro}Introduction}

Recently, it has been discovered that quantum contextuality might
have a significant place in a development quantum communication
\cite{cabello-dambrosio-11,nagata-05}, quantum computation 
\cite{magic-14,bartlett-nature-14}, and lattice theory
\cite{bdm-ndm-mp-fresl-jmp-10,mp-7oa}. This has prompted 
experimental implementation of 4-, 6-, and 8-dimensional contextual
experiments with photons~\cite{simon-zeil00,michler-zeil-00,amselem-cabello-09,liu-09,d-ambrosio-cabello-13,ks-exp-03,canas-cabello-8d-14}, neutrons
\cite{h-rauch06,cabello-fillip-rauch-08,b-rauch-09},
trapped ions \cite{k-cabello-blatt-09},
solid state molecular nuclear spins  \cite{moussa-09}, and paths~\cite{lisonek-14,canas-cabello-14}. 

Experimental contextual tests involve subtle issues, such
as the possibility of noncontextual hidden variable models that
can reproduce quantum mechanical predictions up to arbitrary
precision~\cite{barrett-kent-04}. These models are important because
they show how assignments of predetermined values to dense sets of
projection operators are precluded by any quantum model. Thus,
Spekkens \cite{spekkens-05} introduces generalised noncontextuality
in an attempt to make precise the distinction between classical and
quantum theories, distinguishing the notions of preparation,
transformation, and measurement of noncontextuality and by
doing so demonstrates that even the 2D Hilbert space is not
inherently noncontextual. Kunjwal and Spekkens
\cite{kunjwal-spekkens-15} derive an inequality that does not
assume that the value assignments are deterministic, showing that
noncontextuality cannot be salvaged by abandoning determinism.
Kunjwal \cite{kunjwal-18-arxiv} shows how to compute a
noncontextuality inequality from an invariant derived from a
contextual set/configuration representing an experimental
Kochen-Specker (KS) setup.

This opens up the possibility of finding contextual sets that
provide the best noise robustness in demonstrating contextuality.
The large number of such sets that we show in the present work
can provide a rich source for such an effort.

Quantum contextual configurations that have been elaborated on
the most in the literature are the KS sets, and,
in this paper, we consider just them. In order to obtain KS sets,
so far, various methods of exploiting correlations, symmetries,
geometry, qubit states, Pauli states, Lie algebras, etc., have
been found and used for generating master sets i.e.,~big sets
which contain all smaller contextual sets
\cite{cabell-est-96a,pmmm05a,aravind10,waeg-aravind-jpa-11,mfwap-11,mp-nm-pka-mw-11,waeg-aravind-megill-pavicic-11,waegell-aravind-12,waeg-aravind-pra-13,waeg-aravind-fp-14,waeg-aravind-jpa-15,waeg-aravind-pla-17,pavicic-pra-17}.

All of these methods boil down either to finding a list of vectors
and their $n$-tuples of orthogonalities from which a master set
can be read off or finding a structure, e.g., a polytope,
from which again a list of vectors and orthogonalities can be
read off as well as a master set they build. In the present paper,
we take the simplest possible vector components within an
$n$-dimensional Hilbert space, e.g., $\{0,\pm 1\}$, and via our
algorithms and programs exhaustively build all possible vectors and
their orthogonal $n$-tuples and then filter out KS sets from the
sets in which the vectors are organized. For a particular choice of
components, the chances of getting KS sets are very high. We
generate KS sets for even-dimensional spaces, up to 32, that
properly contain all previously obtained and known KS sets,
present their features and distributions, give examples of
previously unknown sets, and present a blueprint for
implementation of a simple set with a complex coordinatization.


\section{Results}

The main results presented in this paper concern generation of 
contextual sets from several basic vector components. Previous
contextual sets from the literature made use of often complicated
sets of vectors that the authors arrived at, following
particular symmetries, or geometries, or polytope correlations,
or Pauli operators, or qubit states, etc. In contrast, our approach
considers McKay--Megill--Pavi\v ci\'c (MMP) hypergraphs (defined in
Subsection~\ref{subsec:form}) from $n$-dimensional ($n$D)
Hilbert space (${\cal H}^n$, $n\ge 3$) originally consisting of
$n$-tuples (in our approach represented by MMP hypergraph edges) of
orthogonal vectors (MMP hypergraph vertices) which exhaust themselves
in forming configurations/sets of vectors (MMP hypergraphs). Already
in \cite{pmmm05a-corr}, we realised that hypergraphs massively
generated by their non-isomorphic upward construction might satisfy
the Kochen--Specker theorem even when there were no vectors by means
of which they might be represented (see Theorem~\ref{th:ks}), and 
finding coordinatizations for those hypergraphs which might have them,
via standard methods of solving systems of non-linear equations, is
an exponentially complex task solvable only for the smallest
hypergraphs \cite{pmmm05a-corr}. It was, therefore, rather
surprising to us to discover that the hypergraphs formed by very
simple vector components often satisfied the Kochen--Specker
theorem. In this paper, we present a method of generation of KS MMP
hypergraphs, also called KS hypergraphs, via such simple sets of
vector components.

\begin{theorem}\label{th:ks}{\bf (MMP hypergraph reformulation of
    the Kochen--Specker theorem)}
  There are $n${\rm D} {\rm MMP} hypergraphs, i.e., those whose
  each edge contains $n$ vertices, called {\rm KS MMP} hypergraphs,
  to which it is impossible to assign 1s and 0s in such a way that
  \begin{enumerate}
\item[($\alpha$)] No two vertices within any of its edges are both assigned
  the value 1;
\item[($\beta$)] In any of its edges, not all of the vertices are
  assigned the value 0.
\end{enumerate}
\end{theorem}

\begin{figure}
\centering
  \includegraphics[width=0.3\textwidth]{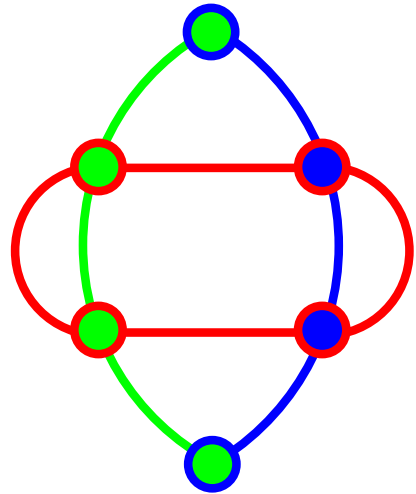}
\caption{The smallest 4D KS MMP hypergraph without a
  coordinatization.}
\label{fig:6-3}
\end{figure}

In Figure~\ref{fig:6-3}, we show the smallest possible 4D KS MMP
hypergraph with six vertices and three edges. We can easily verify
that it is impossible to assign 1 and 0 to its vertices so
as to satisfy the conditions ($\alpha$) and ($\beta$) from
Theorem \ref{th:ks}. For instance, if we assign 1 to the
top green-blue vertex, then, according to the condition
($\alpha$), all of the other vertices contained in the blue and
green edges must be assigned value 0, but, herewith, all four
vertices in the red edge are assigned 0s in violation of
the condition ($\beta$), or, if we assign 1 to the
top red-blue vertex, then, according to the condition
($\alpha$), all the other vertices contained in the blue and
red edges must be assigned value 0, but, herewith, all four
vertices in the green edge are assigned 0s in violation of
the condition ($\beta$). Analogous verifications go through
for the remaining four vertices. We verified that there is
neither a real nor complex vector solution of a corresponding
system of nonlinear equations \cite{pmmm05a-corr}. We have
not tried quaternions as of yet. 

When a coordinatization of a KS MMP hypergraph exists, its
vertices denote $n$-dimensional vectors in ${\cal H}^n$, $n\ge 3$,
and edges designate orthogonal $n$-tuples of vectors containing
the corresponding vertices. In our present approach, a
coordinatization is automatically assigned to each hypergraph by
the very procedure of its generation from the basic vector
components. A KS MMP hypergraph with a given coordinatization
of whatever origin we often simply call a KS {\em set\/}.

\subsection{\label{subsec:form}Formalism}

MMP hypergraphs are those whose edges (of size $n$) intersect each
other in at most $n-2$ vertices~\cite{pmmm05a,pavicic-pra-17}. They are
encoded by means of printable ASCII characters. Vertices are denoted
by one of the following characters: {{\tt 1 2 \dots 9 A B \dots Z a b
\dots z ! " \#} {\$} \% \& ' ( ) * - / : ; \textless\ =
\textgreater\ ? @ [ {$\backslash$} ] \^{} \_ {`} {\{}
{\textbar} \} \textasciitilde} \cite{pmmm05a}. When all of
them are exhausted, one reuses them prefixed by `+',
then again by `++', and so forth. An $n$-dimensional KS set with $k$ vectors
and $m$ $n$-tuples is represented by an MMP hypergraph with
$k$ vertices and $m$ edges which we denote as a $k$-$m$ set.
In its graphical representation, vertices are depicted as dots and
edges as straight or curved lines connecting $m$ orthogonal vertices. 
We handle MMP hypergraphs by means of algorithms in the programs
SHORTD, MMPSTRIP, MMPSUBGRAPH, VECFIND, STATES01, and others
\cite{bdm-ndm-mp-1,pmmm05a-corr,pmm-2-10,bdm-ndm-mp-fresl-jmp-10,mfwap-s-11,mp-nm-pka-mw-11}. In its numerical representation (used for computer
processing), each MMP hypergraph is encoded in a single line in
which all $m$ edges are successively given, separated by commas, and
followed by assignments of coordinatization to $k$ vertices 
(see 18-9 in Subsection~\ref{subsec:vec}). 

\subsection{\label{subsec:vec}KS Vector Lists vs.~Vector
Component MMP Hypergraphs}

In Table \ref{T:1}, we give an overview of most of the $k$-$m$ KS sets
(KS hypergraphs with $m$ vertices and $k$ edges) as defined via lists
and tables of vectors used to build the KS master sets that one can
find in the literature. These master sets serve us to obtain billions
of non-isomorphic smaller KS sets (KS subsets, subhypergraphs) which
define $k$-$m$ {\em classes\/}. In doing so (via the aforementioned
algorithms and programs), we keep to minimal, {\em critical\/}, KS
subhypergraphs in the sense that a removal of any of their edges turns
them into non-KS sets. Critical KS hypergraphs are all we need for an
experimental implementation: additional orthogonalities that bigger
KS sets (containing critical ones) might possess do not add any new
property to the ones that the minimal critical core already has. The
smallest hypergraphs we give in the table are therefore the smallest
criticals. Many more of them, as well as their distributions, the
reader can find in the cited references. Some coordinatizations are
over-complicated in the original literature. For example (as shown in
\cite{pavicic-pra-17}), for the 4D 148-265 master, components 
$\{0,\pm i,\pm 1,\pm\omega,\pm\omega^2\}$, where $\omega=e^{2\pi i/3}$,
suffice for building the coordinatization, and for the  6D 21-7
components $\{0,1,\omega\}$ suffice. In addition, $\{0,\pm 1\}$ suffice for
building the 6D 236-1216.

\begin{table*}
  \caption{\label{T:1}Vector lists from the literature; we call their
    masters {\em list-masters\/}. We shall make use of their vector
    components from the last column to generate master hypergraphs in
    Subsection~\ref{subsec:mas} which we call {\em component-masters\/}.
  $\omega$ is a cubic root of unity: $\omega=e^{2\pi i/3}$.}
\centering
\begin{tabular}{cccccc}
  \textbf{dim} & \textbf{Master Size}
& \textbf{Vector List} & \textbf{List Origin}
&\textbf{Smallest Hypergraph}
   & \textbf{Vector Components}\\
\Xhline{2\arrayrulewidth}
  4D & 24-24 & \cite{peres,kern,cabell-est-96a}
 & \begin{tabular}{@{}c@{}}symmetry,\\geometry\end{tabular}
                     &  {\parbox[c]{1em}{\includegraphics[scale=0.27]{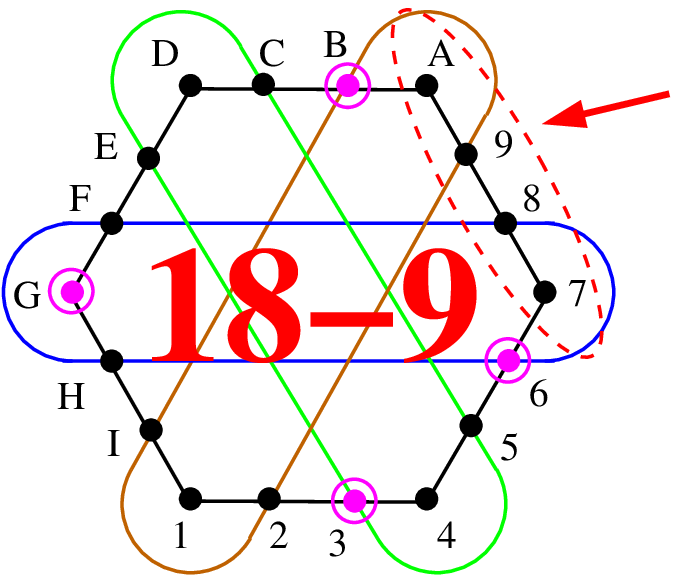}}\hfil}
 & \{0,$\pm 1$\}\\
  4D & 60-105 & \cite{waeg-aravind-jpa-11,pavicic-pra-17} &
 \begin{tabular}{@{}c@{}}Pauli\\operators\end{tabular}
               & {\parbox[c]{1em}{\includegraphics[scale=0.27]{18-9-col-lett.eps}}\hfil}
& \{0,$\pm 1,\pm i$\}\\
  4D & 60-75 & \cite{aravind10,mp-nm-pka-mw-11,mfwap-s-11,pavicic-pra-17} &
\begin{tabular}{@{}c@{}}regular\\polytope\\600-cell\end{tabular}               
& {\parbox[c]{1em}{\includegraphics[scale=0.06]{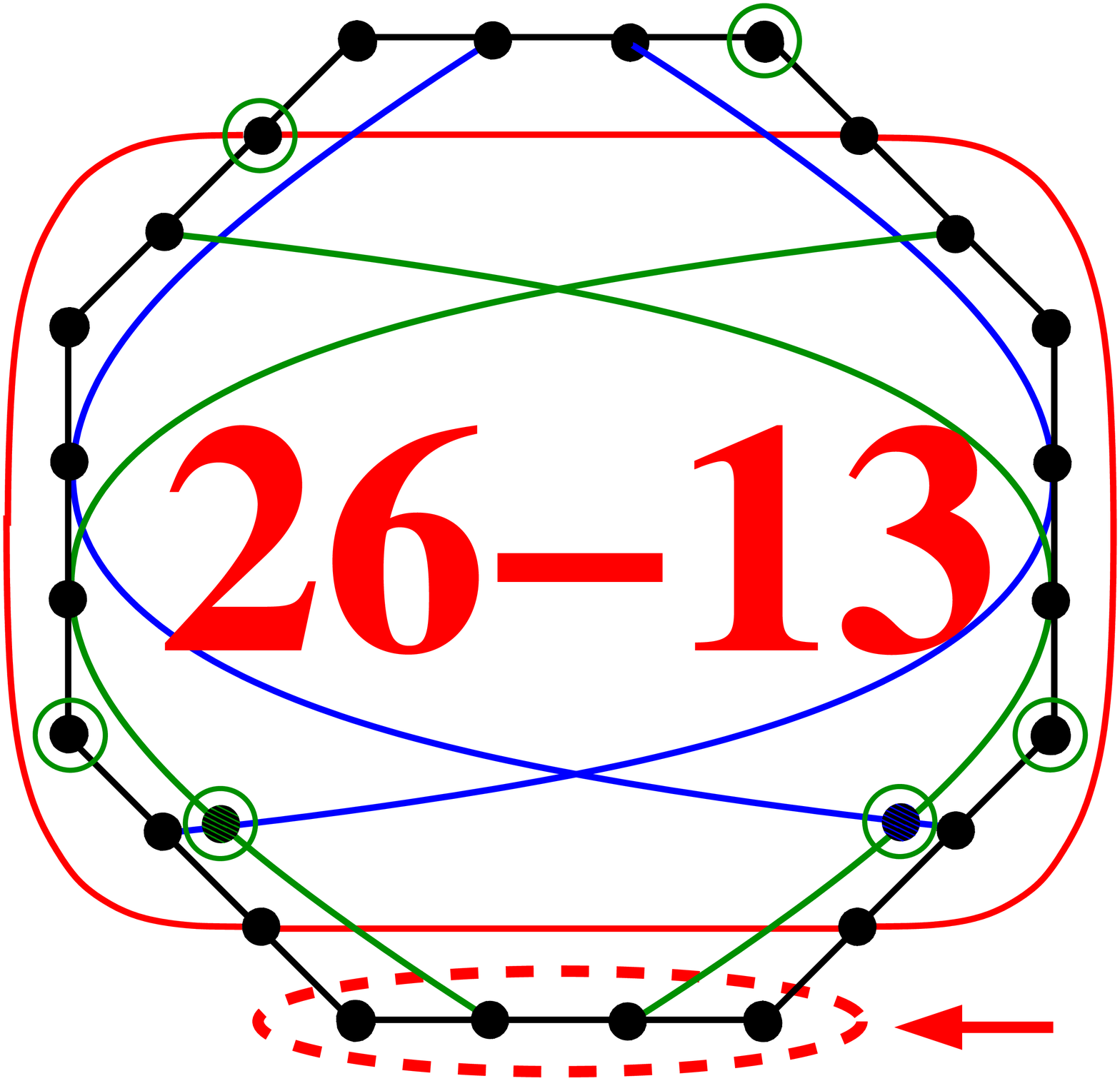}}\hfil}
&\begin{tabular}{@{}c@{}}$\{0,\pm(\sqrt{5}-1)/2,\pm 1,$\\$\pm(\sqrt{5}+1)/2,2\}$\end{tabular}  \\
4D & 148-265 & \cite{waeg-aravind-pla-17,pavicic-pra-17}
& \begin{tabular}{@{}c@{}}Witting\\polytope\end{tabular}               
& {\parbox[c]{4em}{\includegraphics[scale=0.051]{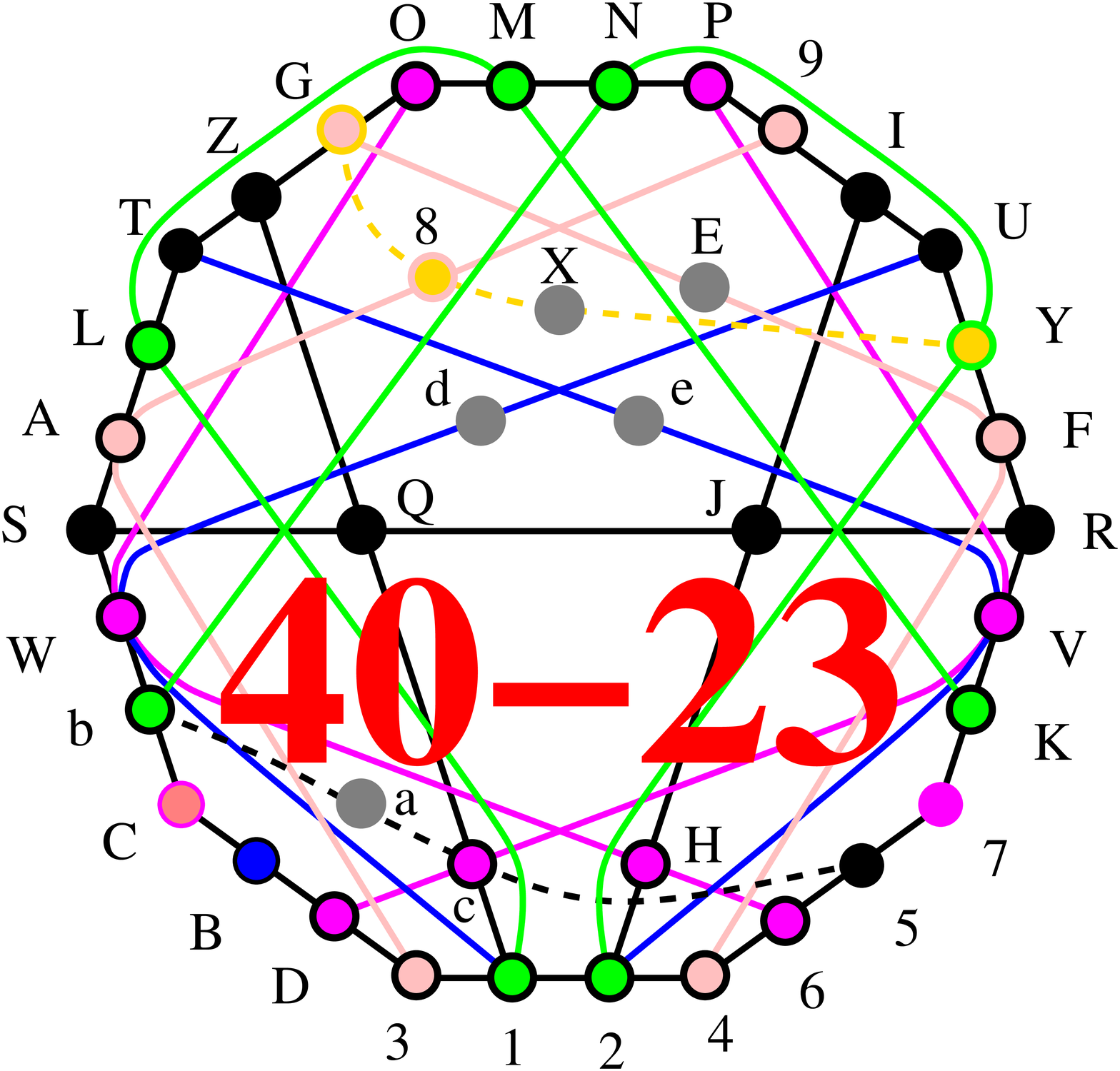}}\hfil}
&\begin{tabular}{@{}c@{}}$\ \ \{0,\pm i,\pm 1,\pm\omega,\pm\omega^2,$\\$\ \pm i\omega^{1/\sqrt{3}},\pm i\omega^{2/\sqrt{3}}\}$\end{tabular} 
  \\
  6D & 21-7 & \cite{lisonek-14} & symmetry 
&{\parbox[c]{2em}{\includegraphics[scale=0.07]{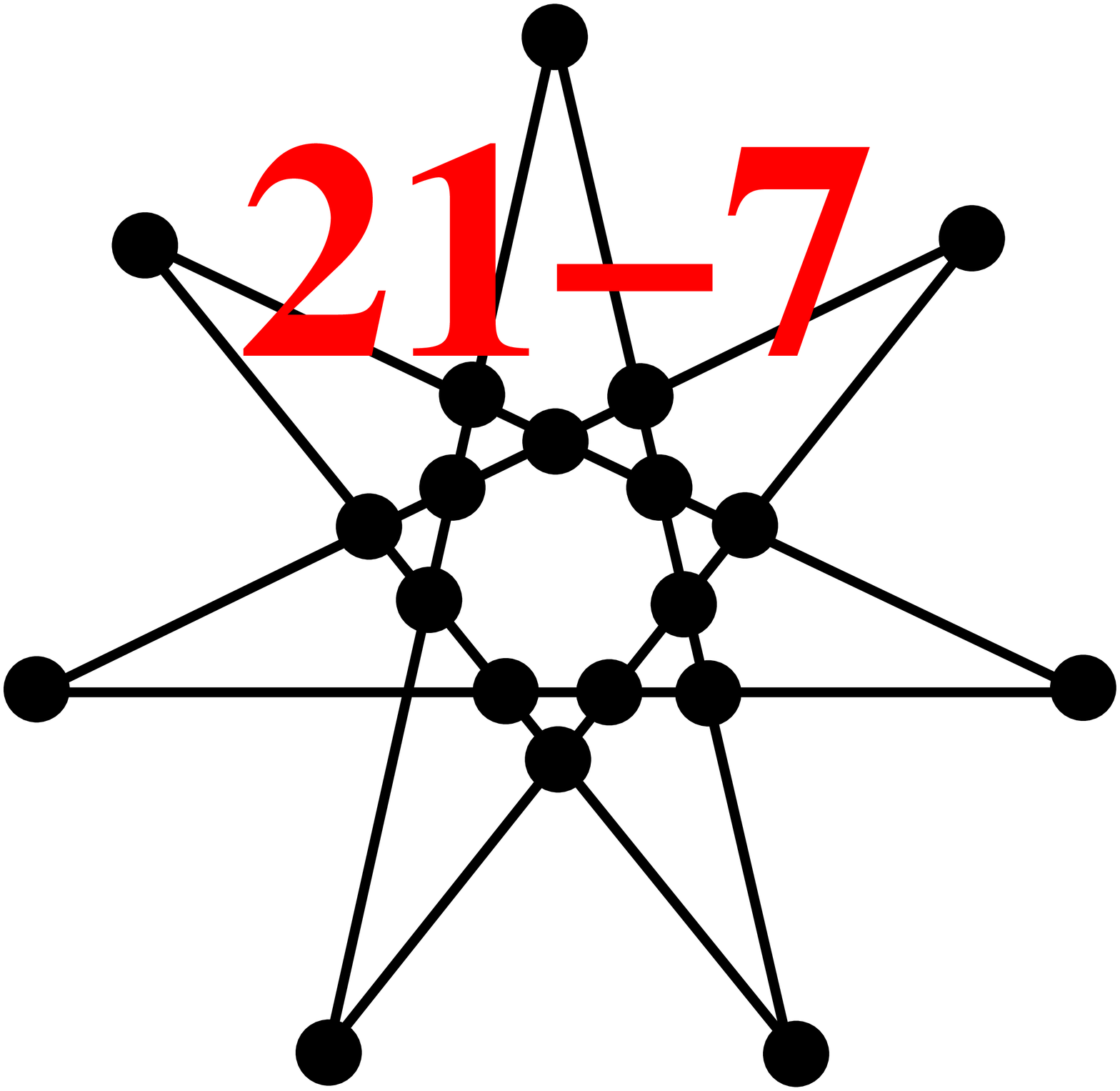}}\hfil}
&  $\{0,1,\omega,\omega^2\}$
  \\
  6D & 236-1216
& \begin{tabular}{@{}c@{}}Aravind \&\\ Waegell\\
2016, \ \cite{pavicic-pra-17}\\\end{tabular}
& \begin{tabular}{@{}c@{}}hypercube\\$\to$hexaract\\Sch\"afli
                         $\{4,3^4\}$\end{tabular}               
&{\parbox[c]{2em}{\includegraphics[scale=0.048]{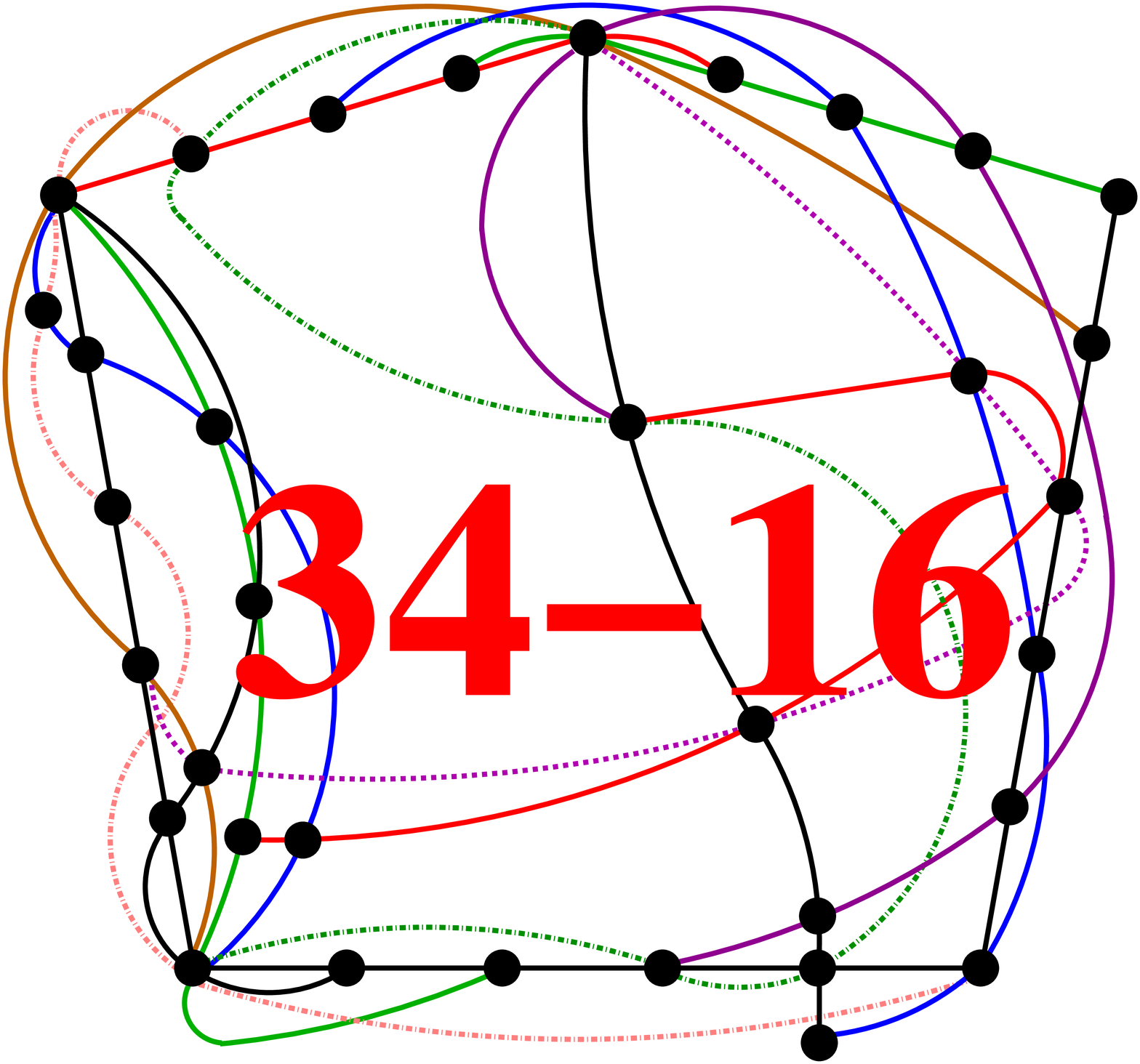}}\hfil}
& \begin{tabular}{@{}c@{}}$\{0,\pm 1/2,\pm 1/\sqrt{3},$\\$\pm 1/\sqrt{2},1\}$\end{tabular} 
  \\
  8D & 36-9 & \cite{pavicic-pra-17} & symmetry               
&{\parbox[c]{2em}{\includegraphics[scale=0.058]{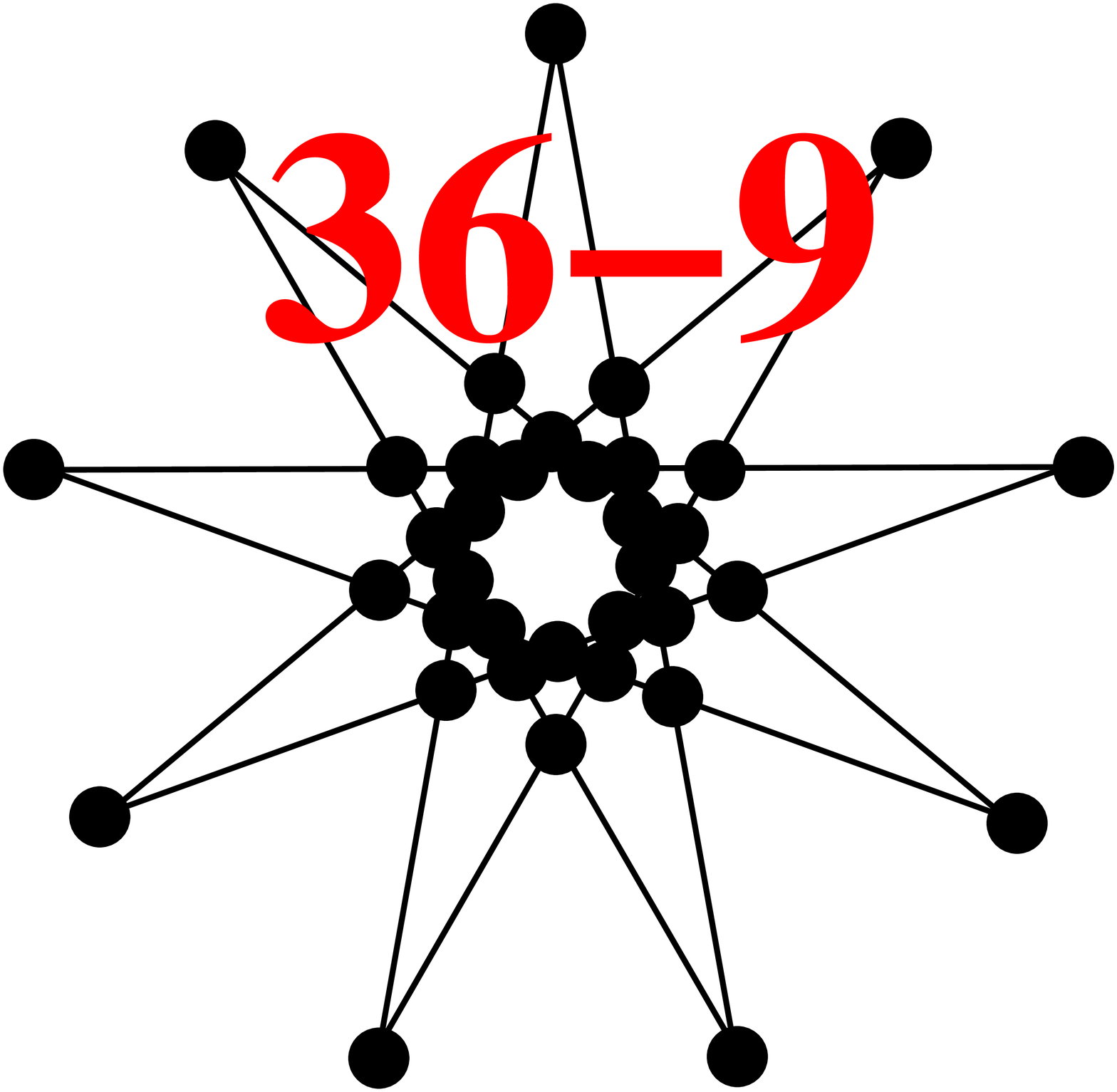}}\hfil}
& $\{0,\pm 1\}$
  \\
  8D & 120-2025 & \cite{waeg-aravind-jpa-15,pavicic-pra-17}
& \begin{tabular}{@{}c@{}}Lie\\algebra\\E8\end{tabular}               
&{\parbox[c]{3em}{\includegraphics[scale=0.082]{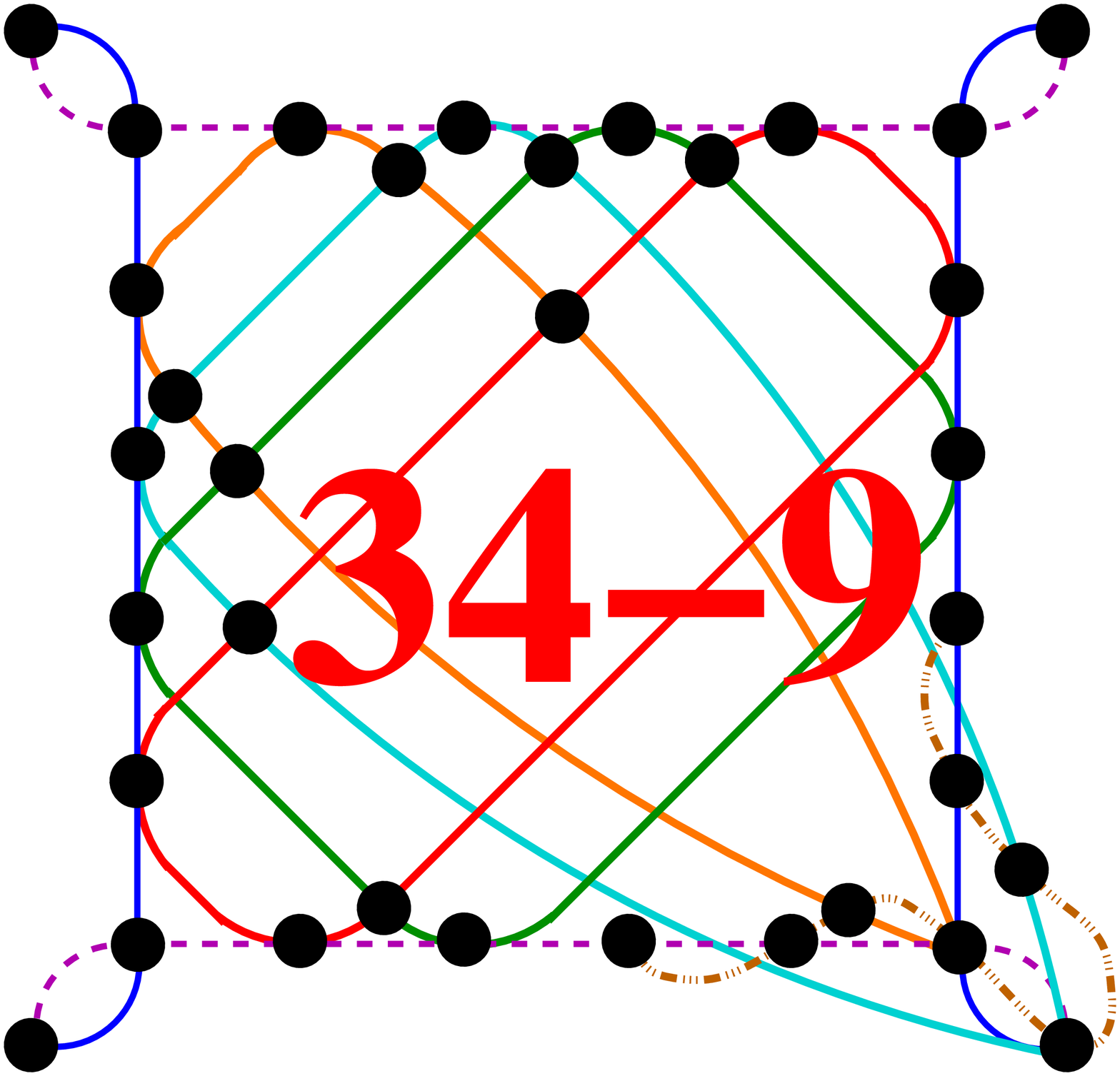}}\hfil}
& \begin{tabular}{@{}c@{}}as given\\in \cite{waeg-aravind-jpa-15}\end{tabular} 
  \\
  16D & 80-265 & \cite{harv-cryss-aravind-12,planat-12,pavicic-pra-17}
& \begin{tabular}{@{}c@{}}Qubit\\states\end{tabular}               
&{\parbox[c]{4em}{\includegraphics[scale=0.11]{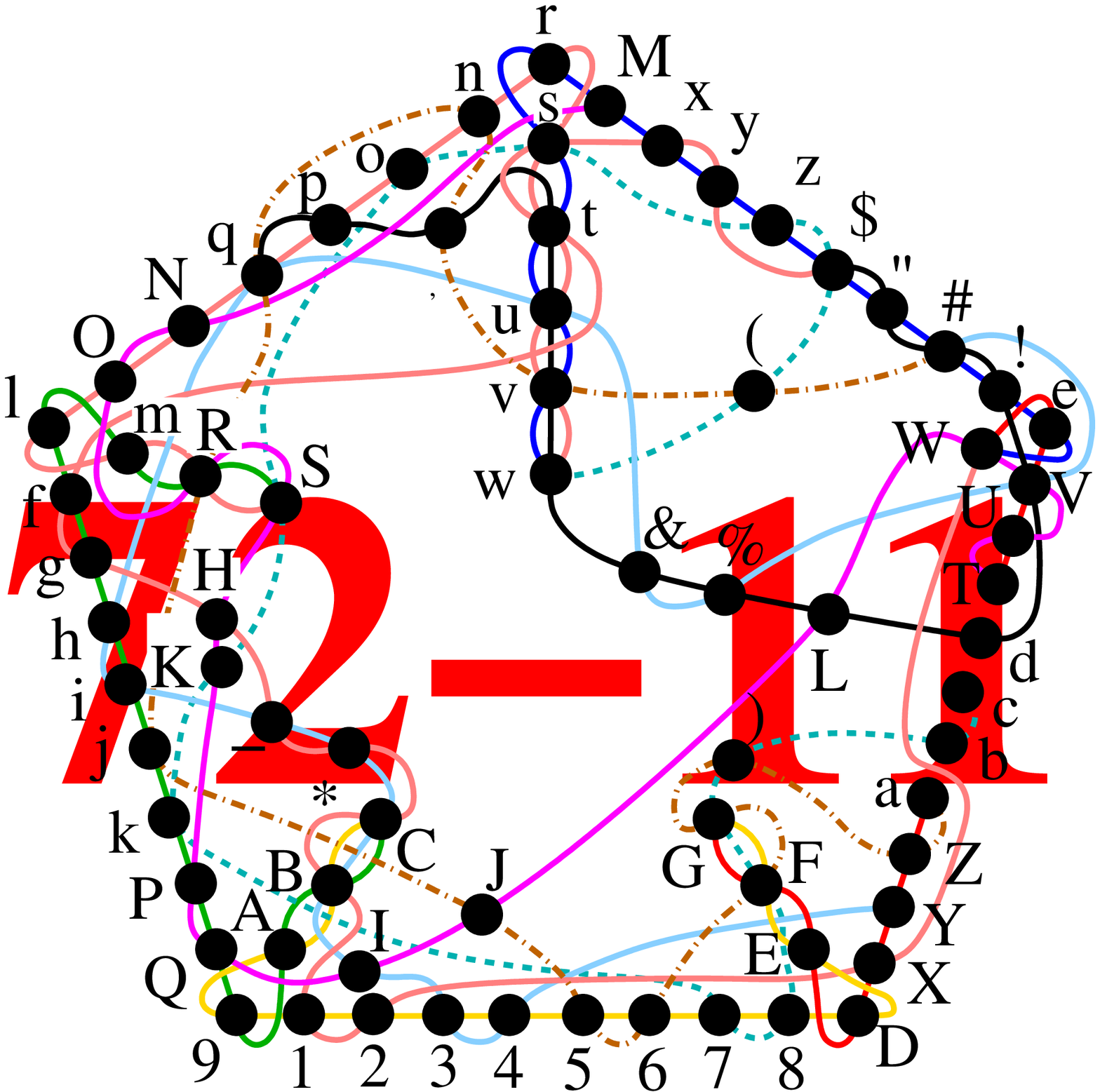}}\hfil}
& $\{0,\pm 1\}$
  \\
  32D & 160-661 & \cite{planat-saniga-12,pavicic-pra-17}
& \begin{tabular}{@{}c@{}}Qubit\\states\end{tabular}               
&{\parbox[c]{5em}{\includegraphics[scale=0.072]{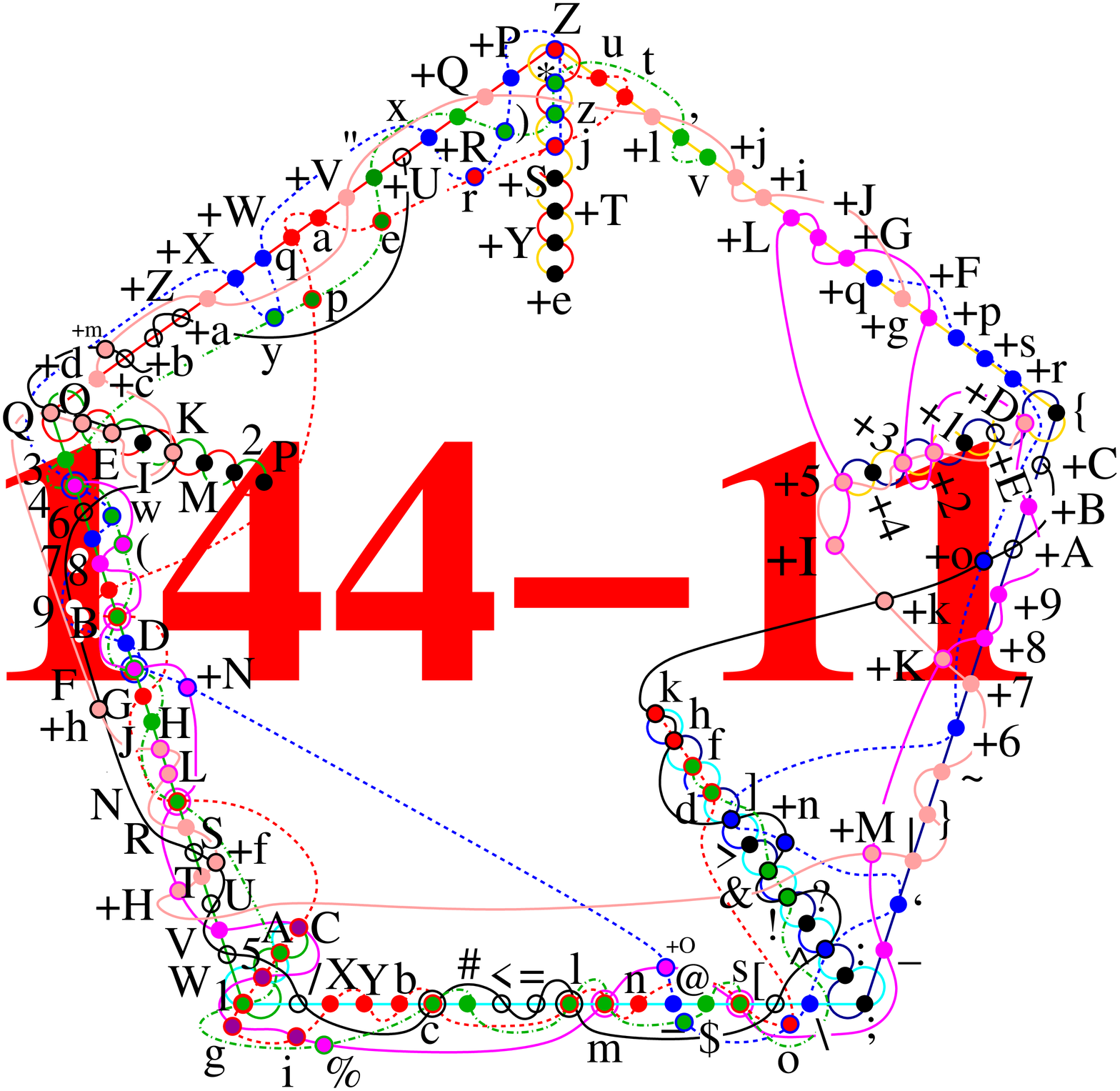}}\hfil}
& $\{0,\pm 1\}$
  \\
\Xhline{3\arrayrulewidth}
\end{tabular}
\end{table*}

Some of the smallest KS hypergraphs in the table have ASCII characters
assigned and some do not. This is to stress that we can assign them
in an arbitrary and random way to any hypergraph and then the
program VECFIND will provide them with a coordinatization in
a fraction of a second. For instance,

\vspace{5pt}
\parindent=0pt
{\bf 18-9}: {\tt 1234,4567,789A,ABCD,DEFG,GHI1,I29B,35CE,\break 68FH.}
{\tt \{1=\{0,0,0,1\},2=\{0,0,1,0\},3=\{1,1,0,0\},4=\{1,\break -1,0,0\},5=\{0,0,1,1\},6=\{1,1,1,-1\},7=\{1,1,-1,1\},\break 8=\{1,-1,1,1\},9=\{1,0,0,-1\},A=\{0,1,1,0\},B=\{1,0,0,\break 1\},C=\{1,-1,1,-1\},D=\{1,1,-1,-1\},E=\{1,-1,-1,1\},F=\break \{0,1,0,1\},G=\{1,0,1,0\},H=\{1,0,-1,0\},I=\{0,1,0,0\}$\!$\}$\!$.}

\vspace{5pt}
(To simplify parsing, this notation delineates vectors with braces
instead of traditional parentheses in order to reserve parentheses
for component expressions.)

\vspace{9pt}
\parindent=20pt
However, a real finding is that we can go the other way round and determine
the KS sets from nothing but vector components $\{0,\pm 1\}$.

\vspace{15pt}

\subsection{\label{subsec:mas}Vector-Component-Generated
    Hypergraph Masters}

\vspace{5pt}
We put simplest possible vector components, which might build vectors
and therefore provide a coordinatization to MMP hypergraphs, into our
program VECFIND. Via its option {\tt -master}, the program builds an
internal list of all possible non-zero vectors containing these
components. From this list, it finds all possible edges of the
hypergraph, which it then generates. MMPSTRIP via its option {\tt -U}
separates unconnected MMP subgraphs. We pipe the obtained hypergraphs
through the program STATES01 to keep those that possess the KS property.
We can use other programs of ours, MMPSTRIP, MMPSHUFFLE, SHORTD,
STATES01, LOOP, etc., to obtain smaller KS subsets and analyze their
features.

\vspace{5pt}
The likelihood that chosen components will give us a KS master
hypergraph and the speed with which it does so depends on
particular features they possess. Here, we will elaborate on some of
them and give a few examples. Features are based on statistics
obtained in the process of generating~hypergraphs:

\begin{enumerate}[\it(i)]
\item the input set of components for generating two-qubit KS
  hypergraphs (4D) should contain number pairs of opposite signs,
  e.g., $\pm 1$, and zero (0); we conjecture that the same holds
  for 3, 4, \dots qubits; with 6D it does not hold literally; e.g.,
  $\{0,1,\omega\}$ generate a KS master; however, the following
  combination of $\omega$'s gives the opposite sign to 1:
  $\omega+\omega^2=-1$;
\item mixing real and complex components gives
  a denser distribution of smaller KS hypergraphs; 
\item reducing the number of components shortens
  the time needed to generate smaller hypergraphs and apparently
  does not affect their distribution. 
\end{enumerate}

\bigskip
Feature {\it(i)} means that, no matter how many different numbers 
we use as our input components, we will not get a KS master
if at least to one of the numbers, the same number with the opposite
sign is not added. Thus, e.g., $\{0,1,-i,2,-3,4,5\}$ or a similar string
does not give any, while $\{0,\pm 1\}$, or $\{0,\pm i\}$, or
$\{0,\pm(\sqrt{5}-1)/2\}$ do. Of course, the latter strings all give
mutually isomorphic KS masters, i.e., one and the same KS master,
if used alone. More specifically, they yield a 40-32 master with 40
vertices and 32 edges as shown in Table \ref{T:2}. When
any of them are used together with other components, although they
would generate different component-masters, all the latter masters
of a particular dimension would have a common smallest hypergraph
as also shown in Table \ref{T:2}.

\begin{table*}
  \caption{\label{T:2} Component-masters we obtained. List-masters are
    given in Table \ref{T:1}. In the last two rows of all but the last
    column, we refer to the result \cite{waeg-aravind-pra-13} that there
    are 16D and 32D criticals with just nine edges. According to the
    conjectured feature {\it(i)\/} above, the masters generated by
    $\{0,\pm 1\}$ should contain those criticals; they did not
    come out in \cite{pavicic-pra-17}, so, we do not know how many
    vertices they have. The smallest ones we obtained are given in
    Table \ref{T:1}. The number of criticals given in the 4th column refer
    to the number of them we successfully generated although there are
    many more of them except in the 40-32 class.} 
\centering
\begin{tabular}{cccccc}
  \Xhline{2\arrayrulewidth}
  \textbf{dim} &\textbf{Vector Components}
  & \begin{tabular}{@{}c@{}}\textbf{Component-Master}\\\textbf{Size}\end{tabular} 
&\begin{tabular}{@{}c@{}}\textbf{\ \ N{\textsuperscript o} of KS Criticals}\\\textbf{\ \ in Master}\end{tabular} 
& \begin{tabular}{@{}c@{}}\textbf{Smallest}\\\textbf{Hypergraph}\end{tabular}
  & \begin{tabular}{@{}c@{}}\textbf{\ \ Contains}\\\textbf{\ \ List-Masters}\end{tabular} 
\\
\Xhline{3\arrayrulewidth}
  4D & \begin{tabular}{@{}c@{}}\{0,$\pm 1$\} \ or \{0,$\pm i$\} \ or\\
      $\{0,\pm(\sqrt{5}-1)/2\}$ or \dots\end{tabular} 
        & 40-32 & 6
& {\parbox[c]{1em}{\includegraphics[scale=0.22]{18-9-col-lett.eps}}\hfil}
& 24-24\\
  4D & \{0,$\pm 1,\pm i$\} & 156-249 & $7.7\times 10^6$
& {\parbox[c]{1em}{\includegraphics[scale=0.22]{18-9-col-lett.eps}}\hfil}
& 24-24, 60-105\\
  4D & \begin{tabular}{@{}c@{}}$\{0,\pm(\sqrt{5}-1)/2,\pm 1,$\\$\pm(\sqrt{5}+1)/2,2\}$\end{tabular} & 2316-3052 & $1.5\times 10^9$
& {\parbox[c]{1em}{\includegraphics[scale=0.22]{18-9-col-lett.eps}}\hfil}
& 24-24, 60-75\\
  4D & $\{0,\pm 1,\pm i,\pm\omega,\pm\omega^2\}$ & 400-1012 
  &$8\times 10^6$       
& {\parbox[c]{1em}{\includegraphics[scale=0.22]{18-9-col-lett.eps}}\hfil}
  &24-24, 60-105, 148-265\\
  6D & $\{0,\pm 1,\omega,\omega^2\}$ & 11808-314446 & $3\times 10^7$
&{\parbox[c]{1em}{\includegraphics[scale=0.05]{21-7-6d-star.eps}}\hfil}
& 21-7, 236-1216
  \\
  8D & $\{0,\pm 1\}$ & 3280-1361376 & $7\times 10^6$
&{\parbox[c]{1em}{\includegraphics[scale=0.05]{34-9-8d.eps}}\hfil}
& 36-9, 120-2025 
  \\
  16D & $\{0,\pm 1\}$ 
 &\begin{tabular}{@{}c@{}}computationally\\too demanding\end{tabular}
               & $4\times 10^6$
  &\begin{tabular}{@{}c@{}}{\ \ \ \parbox[c]{1em}{\includegraphics[scale=0.065]{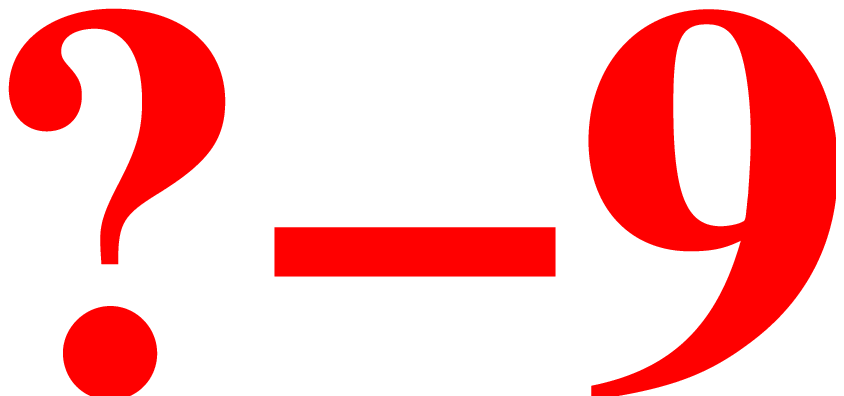}}}\\\qquad\cite{waeg-aravind-pra-13}.\end{tabular}
& 80-265
  \\
  32D & $\{0,\pm 1\}$
  & \begin{tabular}{@{}c@{}}computationally\\too demanding\end{tabular}
  & $2.5\times 10^5$
&\begin{tabular}{@{}c@{}}{\ \ \ \parbox[c]{1em}{\includegraphics[scale=0.065]{9-edges.eps}}}\\\qquad\cite{waeg-aravind-pra-13}.\end{tabular}
& 160-661
  \\
\Xhline{2\arrayrulewidth}
\end{tabular}
\end{table*}

We obtained the following particular results which show the
extent to which component-masters give a more populated
distribution of KS criticals than list-masters. We also closed
several open~questions:
\begin{itemize} 
\item As for the features {\it(ii)} and {\it(iii)} above,
  components $\{0,\pm 1,\omega\}$ generate the master 180-203
  which has the following smallest criticals 18-9, 20\dots 22-11,
  22\dots 26-13, 24\dots 30-15, 30\dots 31-16, 28\dots 35-17,
  33\dots 37-18, etc. This distribution is much denser
  than that of, e.g., the list-master 24-24 with real vectors which
  in the same span of edges consists only of 18-9, 20-11, 22-13,
  and 24-15 criticals or of the list-master 60-75 which starts
  with the 26-13 critical. In Appendix \ref{app:1}, we give a
  detailed description of a 21-11 critical with a complex
  coordinatization and give a blueprint for its experimental
  implementation; 
\item In \cite{lisonek-14}, the reader is challenged to find a
  master set which would contain the "seven context star" 21-7
  KS critical (shown in Tables \ref{T:1} and \ref{T:2}). We find
  that $\{0,1,\omega\}$ generate the 216-153 6D master which
  contains just three criticals 21-7, 27-9, and 33-11,
  $\{0,1,\omega,\omega^2\}$ generate 834-1609 master from
  which we obtained $2.5\times 10^7$ criticals, and
  $\{0,\pm 1,\omega,\omega^2\}$ generate 11808-314446 master from
  which we obtained $3\times 10^7$ criticals, all of them containing
  the seven context star. 27-9 and 39-13 can be viewed as 21-7 with
  a pair of $\delta$-triplets interwoven with 21-7, as shown in
  Figure~\ref{fig:triplets}. The 834-1609 KS master generated from
  $\{0,1,\omega,\omega^2\}$, which were used for a construction of
  21-7 in \cite{lisonek-14}, contains 39-13 as well. Equally so, the
  11808-314446 master.

  \begin{figure}[htp]
\begin{center}
  \includegraphics[width=0.49\textwidth]{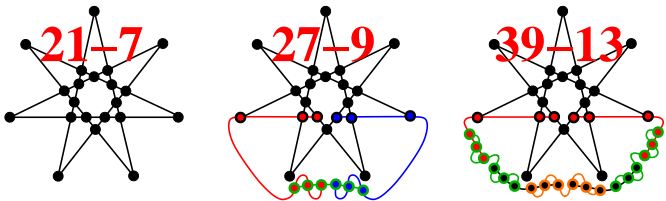}
\end{center}
\caption{21-11 KS set from \cite{lisonek-14} and 27-9 are
  contained in three different master sets, 39-13 in two (together with
  21-11 and 27-9); see the text.}
\label{fig:triplets}
\end{figure}
\item The 60-75 list-master contains criticals with up to 41 edges and
  60 vertices, while the 2316-3052 component-master generated from the
  same vector components contains criticals with up to close to 200
  edges and 300 vertices;
\item The 60-105 list-master contains criticals with up to 40 edges
  and 60 vertices, while the 156-249 component-master generated from
  the same vector components contains criticals with up to at least
  58 edges and 88 vertices;
\item Components $\{0,\pm 1\}$ generate 332-1408 6D master which
  contains the 236-1216 list-master while originally components  
  $\{0,\pm 1/2,\pm 1/\sqrt{3},\pm 1/\sqrt{2},1\}$ were used;
\item In \cite{pavicic-pra-17}, we generated 6D criticals with
  up to 177 vertices and 87 edges from the list-master 236-1216,
  while, now, from the component-master 11808-314446, we obtain
  criticals with up to 201 vertices and 107 edges;
\item We did not generate 16D and 32D masters because that would
  take too many CPU days and we already generated a huge number
  of criticals from submasters which are also defined by means
  of the same vector components in \cite{pavicic-pra-17}.
  See also Section~\ref{sec:met}. 
\end{itemize}

\section{\label{sec:met}Methods}

Our methods for obtaining quantum contextual sets boil down to
algorithms and programs within the MMP language we developed to
generate and handle KS MMP hypergraphs as the most elaborated and
implemented kind of these sets. The programs we make use of,
VECFIND, STATES01, MMPSTRIP, MMPSHUFFLE, SUBGRAPH, LOOP,
SHORTD, etc.,~are freely available from our repository
http://goo.gl/xbx8U2. They are developed in
\cite{bdm-ndm-mp-1,pmmm05a-corr,pm-ql-l-hql2,pmm-2-10,bdm-ndm-mp-fresl-jmp-10,mfwap-11,mp-nm-pka-mw-11,megill-pavicic-mipro-long-17} and extended
for the present elaboration. Each MMP hypergraph can be represented
as a figure for a visualisation but more importantly as a string of
ASCII characters with one line per hypergraph, enabling us to
process millions of them simultaneously by inputting them into
supercomputers and clusters. For the latter elaboration, we
developed other dynamical programs specifically for a supercomputer
or cluster, which enable piping of our files through our programs
in order to parallelize jobs. The programs have the flexibility of
handling practically unlimited number of MMP hypergraph vertices
and edges as we can see from Table \ref{T:2}. The fact that we did
not let our supercomputer run to generate 16D and 36D masters and
our remark that it would be "computationally too demanding" do
not mean that such runs are not feasible with the current computers,
but that they would require too many CPU days on the supercomputer
and that we decided not to burden it with such a task at the present
stage of our research; see the explanation in Subsection~\ref{subsec:mas}.

\section{Conclusions}

The main result we obtain is that our vector component generation
of KS hypergraphs (sets) exhaustively use all possible vectors that
can be constructed from chosen vector components. This is in contrast
to previous studies, which made use of serendipitously obtained lists
of vectors curtailed in number due to various methods applied
to obtain them. Hence, we obtain a thorough and maximally dense
distribution of KS classes in all dimensions whose critical sets
can therefore be much more effectively used for possible implementation
in quantum computation and communication. A comparison of Tables
\ref{T:1} and \ref{T:2} vividly illustrates the difference. 

In Appendix \ref{app:1}, we present a possible experimental
implementation of a KS critical with complex coordinatization
generated from $\{0,\pm 1,\omega\}$. What we immediately notice about
the 21-11 critical from Figure~\ref{fig:21-11} is that the edges are
interwoven in more intricate way than in the 18-9 (which has been
implemented already in several experiments), exhibiting the so-called
$\delta$-feature of the edges forming the biggest loop within a KS
hypergraph. The $\delta$-feature refers to two neighbouring edges
which share two vertices, i.e., intersect each other at two vertices
\cite{pavicic-pra-17}. It stems directly from the representation of
KS configuration with MMP hypergraphs. Notice that the
$\delta$-feature precludes interpretation of practically any KS
hypergraph in an even dimensional Hilbert space by means of 
so-called Greechie diagrams, which by definition require that
two blocks (similar to hypergraph edges) do not share more than
one atom (similar to a vertex) \cite{mp-7oa}, on the one hand,
and that the loops made by the blocks must be of order five or
higher (which is hardly ever realised in even dimensional KS
hypergraphs---see examples in~\cite{pavicic-pra-17}), on the other. 

Our future engagement would be to tackle odd dimensional KS
hypergraphs. Notice that, in a 3D Hilbert space, it is possible to
explore similarities between Greechie diagrams and MMP hypergraphs
because then neither of them can have edges/blocks which share more
than one vertex/atom (via their respective definitions) and loops
in both of them are of the order five or higher
\cite{bdm-ndm-mp-1,pmmm05a}.

\vspace{6pt}

\section*{Author Contributions}

Conceptualization, M.P.; Data Curation, M.P.;
  Formal Analysis, M.P. and N.D.M.; Funding Acquisition, M.P.;
  Investigation, M.P. and N.D.M.; Methodology, M.P. and N.D.M.;
  Project Administration, M.P.; Resources, M.P.;
  Software, M.P. and N.D.M.; Supervision, M.P.;
  Validation, M.P. and N.D.M.; Visualization, M.P.;
  Writing---Original Draft, M.P.;
  Writing---Review and Editing, M.P. and N.D.M.

\begin{acknowledgements}

Supported by the Croatian Science Foundation through project
IP-2014-09-7515, the Ministry of Science and Education (MSE)
of Croatia through the Center of Excellence for Advanced
Materials and Sensing Devices (CEMS) funding, and by 
grants Nos. KK.01.1.1.01.0001 and 533-19-15-0022.
This project was also supported by the Alexander or Humboldt Foundation.
Computational support was provided by the cluster Isabella of
the Zagreb University Computing Centre, by the Croatian National
Grid Infrastructure (CRO-NGI), and by the Center for Advanced
Computing and Modelling (CNRM) for providing computing resources
of the supercomputer Bura at the University of Rijeka in Rijeka,
Croatia. The supercomputer Bura and other information and 
communication technology (ICT) research
infrastructure were acquired through the project
{\it Development of research infrastructure for laboratories of
the University of Rijeka Campus}, which is co-funded by the
European regional development fund. Technical supports of Emir
Imamagi\'c and Daniel Vr\v ci\'c from Isabella and CRO-NGI and
of Miroslav Pu\v skari\'c from CNRM are gratefully acknowledged.
\end{acknowledgements}

\subsection*{Conflicts of Interest}

The authors declare no conflict of interest.

\subsection*{Abbreviations}
The following abbreviations are used in this manuscript:

\noindent 
\begin{tabular}{@{}ll}
KS & Kochen--Specker; defined in Section~\ref{sec:intro}\\
  MMP & McKay-Megill-Pavi\v ci\'c; defined in
        Subsection~\ref{subsec:form}\\
\end{tabular}

\appendix
\section{\label{app:1} 21-11 KS Critical with 
Complex States from ${\cal H}^2\otimes{\cal H}^2$}

\begin{figure}[htp]
\flushleft
  \includegraphics[width=0.49\textwidth]{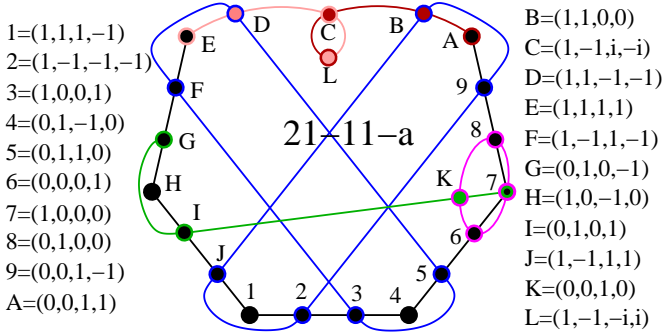}
  \caption{21-11 KS set with a complex coordinatization.}
\label{fig:21-11}
\end{figure}

We present a possible implementation of a KS critical
21-11 with complex coordinatization shown in Fig.~\ref{fig:21-11}.

The vector components of the first qubit on a photon correspond
to a linear (horizontal, $H$, vertical, $V$, diagonal, $D$,
antidiagonal $A$) and circular (right, $R$, left $L$) polarization,
and those of the second qubit to an angular momentum of the photon
$(+2,-2)$ and $(h,v)$. A correspondence between them is
given below. An example of a tensor product of two vectors/states
from ${\cal H}^2\otimes{\cal H}^2$ is:
\begin{align}
|01\rangle&=|0,1\rangle=|0\rangle_1\otimes|1\rangle_2=
\left(
\begin{matrix}
1\\
0\\
\end{matrix}  \right)_1
\otimes\left(
\begin{matrix}
0\\
1\\
\end{matrix}  \right)_2\notag\\
  &=
\left(
\begin{matrix}
1    \left(
     \begin{matrix}
      0\\
      1\\
    \end{matrix}  
    \right)
\\
\\
0    \left(
     \begin{matrix}
      0\\
      1\\
    \end{matrix}  
    \right)
\\
\end{matrix}  \right)
=\left(
\begin{matrix}
     0\\
     1\\
     0\\
     0\\
\end{matrix}  \right).\hfill\phantom{.}
\end{align}

This is our vector {\tt 8} from Figure~\ref{fig:21-11}. Since we
are interested in the qubit states, we are going to proceed in
reverse---from 4-vectors to tensor products of polarization and
angular momentum states. Let us first define them:

\begin{align}
&|H\rangle=\left(
     \begin{matrix}
      1\\
      0\\
    \end{matrix}  
  \right)_1;\quad
|V\rangle=\left(
     \begin{matrix}
      0\\
      1\\
    \end{matrix}  
    \right)_1;\quad  
|D\rangle=\frac{1}{\sqrt{2}}\left(
     \begin{matrix}
      1\\
      1\\
    \end{matrix}  
    \right)_1;\quad  \notag\\
&|A\rangle=\frac{1}{\sqrt{2}}\left(
     \begin{matrix}
      -1\\
      1\\
    \end{matrix}  
    \right)_1;\   
|R\rangle=\frac{1}{\sqrt{2}}\left(
     \begin{matrix}
      1\\
      i\\
    \end{matrix}  
    \right)_1;\ 
|L\rangle=\frac{1}{\sqrt{2}}\left(
     \begin{matrix}
      1\\
      -i\\
    \end{matrix}  
    \right)_1;\notag\\    
 &|+2\rangle=\left(
     \begin{matrix}
      1\\
      0\\
    \end{matrix}  
  \right)_2;\quad
|-2\rangle=\left(
     \begin{matrix}
      0\\
      1\\
    \end{matrix}  
    \right)_2;\quad  
|h\rangle=\frac{1}{\sqrt{2}}\left(
     \begin{matrix}
      1\\
      1\\
    \end{matrix}  
    \right)_2;\quad \quad\notag\\   
&|v\rangle=\frac{1}{\sqrt{2}}\left(
     \begin{matrix}
      1\\
      -1\\
    \end{matrix}  
  \right)_2.
\end{align}
  
Now, one can read off our vertex states as follows:
\begin{align}
&{\tt 1}=\left(
\begin{matrix}
     1\\
     1\\
     1\\
     -1\\
   \end{matrix}  \right)
\to
\frac{1}{2}
  \left(
\begin{matrix}
     1\\
     1\\
     1\\
     -1\\
   \end{matrix}  \right)
=
  \frac{1}{2}(\left(
\begin{matrix}
     1\\
     1\\
     0\\
     0\\
   \end{matrix}  \right)+
  \left(
\begin{matrix}
     0\\
     0\\
     1\\
     -1\\
\end{matrix}  \right) )\notag\\
&=\frac{1}{\sqrt{2}}(\frac{1}{\sqrt{2}}\left(
\begin{matrix}
1    \left(
     \begin{matrix}
      1\\
      1\\
    \end{matrix}  
    \right)
\\
\\
0    \left(
     \begin{matrix}
      1\\
      1\\
    \end{matrix}  
    \right)
\\
\end{matrix}  \right)+
\frac{1}{\sqrt{2}}\left(
\begin{matrix}
0    \left(
     \begin{matrix}
      1\\
      -1\\
    \end{matrix}  
    \right)
\\
\\
1    \left(
     \begin{matrix}
      1\\
      -1\\
    \end{matrix}  
    \right)
\\
\end{matrix}  \right))\notag\\
 & =
\frac{1}{\sqrt{2}}(\left(
\begin{matrix}
1\\
0\\
\end{matrix}  \right)_1
\otimes \frac{1}{\sqrt{2}}\left(
\begin{matrix}
1\\
1\\
\end{matrix}  \right)_2
+  \left(
\begin{matrix}
0\\
1\\
\end{matrix}  \right)_1
\otimes\frac{1}{\sqrt{2}}\left(
\begin{matrix}
1\\
-1\\
\end{matrix}  \right)_2\notag\\
&=\frac{1}{\sqrt{2}}(|H\rangle|h\rangle+|V\rangle|v\rangle)=
\frac{1}{\sqrt{2}}(|D\rangle|+2\rangle-|A\rangle|-2\rangle).
\end{align}

\begin{align}
&{\tt 2}=\left(
\begin{matrix}
     1\\
     -1\\
     -1\\
     -1\\
   \end{matrix}  \right)
\to
 \frac{1}{2} \left(
\begin{matrix}
     1\\
     -1\\
     -1\\
     -1\\
   \end{matrix}  \right)
=\frac{1}{2}(\left(
\begin{matrix}
     1\\
     -1\\
     0\\
     0\\
   \end{matrix}  \right)-
\left(
\begin{matrix}
     0\\
     0\\
     1\\
     1\\
\end{matrix}  \right))\notag\\
&=\frac{1}{\sqrt{2}}(\frac{1}{\sqrt{2}}\left(
\begin{matrix}
1    \left(
     \begin{matrix}
      1\\
      -1\\
    \end{matrix}  
    \right)
\\
\\
0    \left(
     \begin{matrix}
      1\\
      -1\\
    \end{matrix}  
    \right)
\\
\end{matrix}  \right)-
\frac{1}{\sqrt{2}}\left(
\begin{matrix}
0    \left(
     \begin{matrix}
      1\\
      1\\
    \end{matrix}  
    \right)
\\
\\
1    \left(
     \begin{matrix}
      1\\
      1\\
    \end{matrix}  
    \right)
\\
\end{matrix}  \right))\notag\\
&=\frac{1}{\sqrt{2}}(
  \left(
\begin{matrix}
1\\
0\\
\end{matrix}  \right)_1
\otimes\frac{1}{\sqrt{2}}\left(
\begin{matrix}
1\\
-1\\
\end{matrix}  \right)_2
+  \left(
\begin{matrix}
0\\
1\\
\end{matrix}  \right)_1
\otimes\frac{1}{\sqrt{2}}\left(
\begin{matrix}
1\\
1\\
\end{matrix}  \right)_2)\notag\\
&=\frac{1}{\sqrt{2}}(|H\rangle|v\rangle-|V\rangle|h\rangle)=
-\frac{1}{\sqrt{2}}(|D\rangle|-2\rangle+|A\rangle|+2\rangle)
  \label{eq:ver2}
\end{align}

\begin{align}
  &{\tt 3}=
  \left(
\begin{matrix}
     1\\
     0\\
     0\\
     1\\
   \end{matrix}  \right)
\to  \frac{1}{\sqrt{2}}\left(
\begin{matrix}
     1\\
     0\\
     0\\
     1\\
   \end{matrix}  \right)
=\frac{1}{\sqrt{2}}(\left(
\begin{matrix}
     1\\
     0\\
     0\\
     0\\
   \end{matrix}  \right)+
  \left(
\begin{matrix}
     0\\
     0\\
     0\\
     1\\
\end{matrix}  \right))\notag\\
&=\frac{1}{\sqrt{2}}(\left(
\begin{matrix}
1    \left(
     \begin{matrix}
      1\\
      0\\
    \end{matrix}  
    \right)
\\
\\
0    \left(
     \begin{matrix}
      1\\
      0\\
    \end{matrix}  
    \right)
\\
\end{matrix}  \right)+
\left(
\begin{matrix}
0    \left(
     \begin{matrix}
      0\\
      1\\
    \end{matrix}  
    \right)
\\
\\
1    \left(
     \begin{matrix}
      0\\
      1\\
    \end{matrix}  
    \right)
\\
\end{matrix}  \right))\notag\\
& =
 \frac{1}{\sqrt{2}}( \left(
\begin{matrix}
1\\
0\\
\end{matrix}  \right)_1
\otimes\left(
\begin{matrix}
1\\
0\\
\end{matrix}  \right)_2
+  \left(
\begin{matrix}
0\\
1\\
\end{matrix}  \right)_1
\otimes\left(
\begin{matrix}
0\\
1\\
\end{matrix}  \right)_2)\notag\\
&=\frac{1}{\sqrt{2}}(|H\rangle|+2\rangle+|V\rangle|-2\rangle)
  \label{eq:ver3}
\end{align}

\begin{align}
&{\tt 4}=\left(
\begin{matrix}
     0\\
     1\\
     -1\\
     0\\
   \end{matrix}  \right)
\to\frac{1}{\sqrt{2}}(\left(
\begin{matrix}
     0\\
     1\\
     0\\
     0\\
   \end{matrix}  \right)-
  \left(
\begin{matrix}
     0\\
     0\\
     1\\
     0\\
\end{matrix}  \right))\notag\\
&=\frac{1}{\sqrt{2}}(\left(
\begin{matrix}
1    \left(
     \begin{matrix}
      0\\
      1\\
    \end{matrix}  
    \right)
\\
\\
0    \left(
     \begin{matrix}
      0\\
      1\\
    \end{matrix}  
    \right)
\\
\end{matrix}  \right)-
\left(
\begin{matrix}
0    \left(
     \begin{matrix}
      1\\
      0\\
    \end{matrix}  
    \right)
\\
\\
1    \left(
     \begin{matrix}
      1\\
      0\\
    \end{matrix}  
    \right)
\\
\end{matrix}  \right))\notag\\
  &=\frac{
  \left(
\begin{matrix}
1\\
0\\
\end{matrix}  \right)_1
\otimes\left(
\begin{matrix}
0\\
1\\
\end{matrix}  \right)_2
-  \left(
\begin{matrix}
0\\
1\\
\end{matrix}  \right)_1
\otimes\left(
\begin{matrix}
1\\
0\\
\end{matrix}  \right)_2}{\sqrt{2}}\notag\\
  &=\frac{1}{\sqrt{2}}(|H\rangle|-2\rangle-|V\rangle|+2\rangle)
  \label{eq:ver4}
\end{align}

\begin{align}
&{\tt 5}=\left(
\begin{matrix}
     0\\
     1\\
     1\\
     0\\
   \end{matrix}  \right)
\to
\frac{1}{\sqrt{2}}(\left(
\begin{matrix}
1    \left(
     \begin{matrix}
      0\\
      1\\
    \end{matrix}  
    \right)
\\
\\
0    \left(
     \begin{matrix}
      0\\
      1\\
    \end{matrix}  
    \right)
\\
\end{matrix}  \right)+
\left(
\begin{matrix}
0    \left(
     \begin{matrix}
      1\\
      0\\
    \end{matrix}  
    \right)
\\
\\
1    \left(
     \begin{matrix}
      1\\
      0\\
    \end{matrix}  
    \right)
\\
\end{matrix}  \right))\notag\\
&=\frac{1}{\sqrt{2}}(
  \left(
\begin{matrix}
1\\
0\\
\end{matrix}  \right)_1
\otimes\left(
\begin{matrix}
0\\
1\\
\end{matrix}  \right)_2
+  \left(
\begin{matrix}
0\\
1\\
\end{matrix}  \right)_1
\otimes\left(
\begin{matrix}
1\\
0\\
\end{matrix}  \right)_2)\notag\\
  &=\frac{1}{\sqrt{2}}(|H\rangle|-2\rangle+|V\rangle|+2\rangle)
  \label{eq:ver5}
\end{align}

\begin{align}
{\tt 6}=
\left(
\begin{matrix}
     0\\
     0\\
     0\\
     1\\
   \end{matrix}  \right)
\to
\left(\begin{matrix}
0    \left(
     \begin{matrix}
      0\\
      1\\
    \end{matrix}  
    \right)
\\
\\
1    \left(
     \begin{matrix}
      0\\
      1\\
    \end{matrix}  
    \right)
\\
\end{matrix}  \right)
 =
  \left(
\begin{matrix}
0\\
1\\
\end{matrix}  \right)_1
\otimes\left(
\begin{matrix}
0\\
1\\
\end{matrix}  \right)_2
=|V\rangle|-2\rangle
  \label{eq:ver6}
\end{align}

\begin{align}
  {\tt 7}=
\left(
\begin{matrix}
     1\\
     0\\
     0\\
     0\\
   \end{matrix}  \right)
\to
\left(\begin{matrix}
1    \left(
     \begin{matrix}
      1\\
      0\\
    \end{matrix}  
    \right)
\\
\\
0    \left(
     \begin{matrix}
      1\\
      0\\
    \end{matrix}  
    \right)
\\
\end{matrix}  \right)
 =
  \left(
\begin{matrix}
1\\
0\\
\end{matrix}  \right)_1
\otimes\left(
\begin{matrix}
1\\
0\\
\end{matrix}  \right)_2
=|H\rangle|+2\rangle
  \label{eq:ver7}
\end{align}

\begin{align}
  {\tt 8}=
\left(
\begin{matrix}
     0\\
     1\\
     0\\
     0\\
   \end{matrix}  \right)
\to
\left(\begin{matrix}
1    \left(
     \begin{matrix}
      0\\
      1\\
    \end{matrix}  
    \right)
\\
\\
0    \left(
     \begin{matrix}
      0\\
      1\\
    \end{matrix}  
    \right)
\\
\end{matrix}  \right)
 =
  \left(
\begin{matrix}
1\\
0\\
\end{matrix}  \right)_1
\otimes\left(
\begin{matrix}
0\\
1\\
\end{matrix}  \right)_2
=|H\rangle|-2\rangle
  \label{eq:ver8}
\end{align}

\begin{align}
&{\tt 9}=\left(
\begin{matrix}
     0\\
     0\\
     1\\
     -1\\
   \end{matrix}  \right)
\to
\frac{1}{\sqrt{2}}\left(
\begin{matrix}
0    \left(
     \begin{matrix}
      1\\
      -1\\
    \end{matrix}  
    \right)
\\
\\
1    \left(
     \begin{matrix}
      1\\
      -1\\
    \end{matrix}  
    \right)
\\
\end{matrix}  \right)\notag\\
  &=\frac{1}{\sqrt{2}}
  \left(
\begin{matrix}
0\\
1\\
\end{matrix}  \right)_1
\otimes\left(
\begin{matrix}
1\\
-1\\
\end{matrix}  \right)_2
  =|V\rangle|v\rangle
  \label{eq:ver9}
\end{align}

\begin{align}
{\tt A}=\left(
\begin{matrix}
     0\\
     0\\
     1\\
     1\\
   \end{matrix}  \right)
\to
\frac{1}{\sqrt{2}}\left(
\begin{matrix}
0    \left(
     \begin{matrix}
      1\\
      1\\
    \end{matrix}  
    \right)
\\
\\
1    \left(
     \begin{matrix}
      1\\
      1\\
    \end{matrix}  
    \right)
\\
\end{matrix}  \right)
  =\frac{1}{\sqrt{2}}
  \left(
\begin{matrix}
0\\
1\\
\end{matrix}  \right)_1
\otimes\left(
\begin{matrix}
1\\
1\\
\end{matrix}  \right)_2
  =|V\rangle|h\rangle
  \label{eq:verA}
\end{align}

\begin{align}
{\tt B}=\left(
\begin{matrix}
     1\\
     1\\
     0\\
     0\\
   \end{matrix}  \right)
\to
\frac{1}{\sqrt{2}}\left(
\begin{matrix}
1    \left(
     \begin{matrix}
      1\\
      1\\
    \end{matrix}  
    \right)
\\
\\
0    \left(
     \begin{matrix}
      1\\
      1\\
    \end{matrix}  
    \right)
\\
\end{matrix}  \right)
  =\frac{1}{\sqrt{2}}
  \left(
\begin{matrix}
1\\
0\\
\end{matrix}  \right)_1
\otimes\left(
\begin{matrix}
1\\
1\\
\end{matrix}  \right)_2
  =|H\rangle|h\rangle
  \label{eq:verB}
\end{align}

\begin{align}
&{\tt C}=\left(
\begin{matrix}
     1\\
     -1\\
     i\\
     -i\\
   \end{matrix}  \right)
\to
\frac{1}{2}\left(
\begin{matrix}
1    \left(
     \begin{matrix}
      1\\
      -1\\
    \end{matrix}  
    \right)
\\
\\
i   \left(
     \begin{matrix}
      1\\
      -1\\
    \end{matrix}  
    \right)
\\
\end{matrix}  \right)\notag\\
  &=\frac{1}{\sqrt{2}}
  \left(
\begin{matrix}
1\\
i\\
\end{matrix}  \right)_1
\otimes\frac{1}{\sqrt{2}}\left(
\begin{matrix}
1\\
-1\\
\end{matrix}  \right)_2
  =|R\rangle|v\rangle
  \label{eq:verC}
\end{align}

\begin{align}
&{\tt D}=\left(
\begin{matrix}
     1\\
     1\\
     -1\\
     -1\\
   \end{matrix}  \right)
\to
\frac{1}{2}\left(
\begin{matrix}
1    \left(
     \begin{matrix}
      1\\
      1\\
    \end{matrix}  
    \right)
\\
\\
-1   \left(
     \begin{matrix}
      1\\
      1\\
    \end{matrix}  
    \right)
\\
\end{matrix}  \right)\notag\\
  &=\frac{1}{\sqrt{2}}
  \left(
\begin{matrix}
1\\
-1\\
\end{matrix}  \right)_1
\otimes\frac{1}{\sqrt{2}}\left(
\begin{matrix}
1\\
1\\
\end{matrix}  \right)_2
  =-|A\rangle|h\rangle
  \label{eq:verD}
\end{align}

\begin{align}
&{\tt E}=\left(
\begin{matrix}
     1\\
     1\\
     1\\
     1\\
   \end{matrix}  \right)
\to
\frac{1}{2}\left(
\begin{matrix}
1    \left(
     \begin{matrix}
      1\\
      1\\
    \end{matrix}  
    \right)
\\
\\
1   \left(
     \begin{matrix}
      1\\
      1\\
    \end{matrix}  
    \right)
\\
\end{matrix}  \right)\notag\\
  &=\frac{1}{\sqrt{2}}
  \left(
\begin{matrix}
1\\
1\\
\end{matrix}  \right)_1
\otimes\frac{1}{\sqrt{2}}\left(
\begin{matrix}
1\\
1\\
\end{matrix}  \right)_2
  =|D\rangle|h\rangle
  \label{eq:verE}
\end{align}

\begin{align}
&{\tt F}=\left(
\begin{matrix}
     1\\
     -1\\
     1\\
     -1\\
   \end{matrix}  \right)
\to
\frac{1}{2}\left(
\begin{matrix}
1    \left(
     \begin{matrix}
      1\\
      -1\\
    \end{matrix}  
    \right)
\\
\\
1   \left(
     \begin{matrix}
      1\\
      -1\\
    \end{matrix}  
    \right)
\\
\end{matrix}  \right)\notag\\
  &=\frac{1}{\sqrt{2}}
  \left(
\begin{matrix}
1\\
1\\
\end{matrix}  \right)_1
\otimes\frac{1}{\sqrt{2}}\left(
\begin{matrix}
1\\
-1\\
\end{matrix}  \right)_2
  =|D\rangle|v\rangle
  \label{eq:verF}
\end{align}

\begin{align}
&{\tt G}=\left(
\begin{matrix}
     0\\
     1\\
     0\\
     -1\\
   \end{matrix}  \right)
\to
\frac{1}{\sqrt{2}}\left(
\begin{matrix}
1    \left(
     \begin{matrix}
      0\\
      1\\
    \end{matrix}  
    \right)
\\
\\
-1   \left(
     \begin{matrix}
      0\\
      1\\
    \end{matrix}  
    \right)
\\
\end{matrix}  \right)\notag\\
  &=\frac{1}{\sqrt{2}}
  \left(
\begin{matrix}
1\\
-1\\
\end{matrix}  \right)_1
\otimes\left(
\begin{matrix}
0\\
1\\
\end{matrix}  \right)_2
  =-|A\rangle|-2\rangle
  \label{eq:verG}
\end{align}

\begin{align}
&{\tt H}=\left(
\begin{matrix}
     1\\
     0\\
     -1\\
     0\\
   \end{matrix}  \right)
\to
\frac{1}{\sqrt{2}}\left(
\begin{matrix}
1    \left(
     \begin{matrix}
      1\\
      0\\
    \end{matrix}  
    \right)
\\
\\
-1   \left(
     \begin{matrix}
      1\\
      0\\
    \end{matrix}  
    \right)
\\
\end{matrix}  \right)\notag\\
  &=\frac{1}{\sqrt{2}}
  \left(
\begin{matrix}
1\\
-1\\
\end{matrix}  \right)_1
\otimes\left(
\begin{matrix}
1\\
0\\
\end{matrix}  \right)_2
  =-|A\rangle|+2\rangle
  \label{eq:verH}
\end{align}

\begin{align}
&{\tt I}=\left(
\begin{matrix}
     0\\
     1\\
     0\\
     1\\
   \end{matrix}  \right)
\to
\frac{1}{\sqrt{2}}\left(
\begin{matrix}
1    \left(
     \begin{matrix}
      1\\
      0\\
    \end{matrix}  
    \right)
\\
\\
-1   \left(
     \begin{matrix}
      1\\
      0\\
    \end{matrix}  
    \right)
\\
\end{matrix}  \right)\notag\\
  &=\frac{1}{\sqrt{2}}
  \left(
\begin{matrix}
1\\
-1\\
\end{matrix}  \right)_1
\otimes\left(
\begin{matrix}
1\\
0\\
\end{matrix}  \right)_2
  =|D\rangle|-2\rangle
  \label{eq:verI}
\end{align}

\begin{align}
&{\tt J}=\left(
\begin{matrix}
     1\\
     -1\\
     1\\
     1\\
   \end{matrix}  \right)
\to
=\frac{1}{2}(\left(
\begin{matrix}
1    \left(
     \begin{matrix}
      1\\
      0\\
    \end{matrix}  
    \right)
\\
\\
1    \left(
     \begin{matrix}
      1\\
      0\\
    \end{matrix}  
    \right)
\\
\end{matrix}  \right)+
\left(
\begin{matrix}
-1    \left(
     \begin{matrix}
      0\\
      1\\
    \end{matrix}  
    \right)
\\
\\
1    \left(
     \begin{matrix}
      0\\
      1\\
    \end{matrix}  
    \right)
\\
\end{matrix}  \right))\notag\\
&=\frac{1}{\sqrt{2}}(
  \frac{1}{\sqrt{2}}\left(
\begin{matrix}
1\\
1\\
\end{matrix}  \right)_1
\otimes\left(
\begin{matrix}
0\\
1\\
\end{matrix}  \right)_2
+ \frac{1}{\sqrt{2}} \left(
\begin{matrix}
-1\\
1\\
\end{matrix}  \right)_1
\otimes\left(
\begin{matrix}
0\\
1\\
\end{matrix}  \right)_2)\notag\\
&=\frac{1}{\sqrt{2}}(|D\rangle|+2\rangle+|A\rangle|-2\rangle)
  \label{eq:verJ}
\end{align}

\begin{align}
  &{\tt K}=
\left(
\begin{matrix}
     0\\
     0\\
     1\\
     0\\
   \end{matrix}  \right)
\to
\left(\begin{matrix}
0    \left(
     \begin{matrix}
      1\\
      0\\
    \end{matrix}  
    \right)
\\
\\
1    \left(
     \begin{matrix}
      1\\
      0\\
    \end{matrix}  
    \right)
\\
\end{matrix}  \right)\notag\\
 &=
  \left(
\begin{matrix}
0\\
1\\
\end{matrix}  \right)_1
\otimes\left(
\begin{matrix}
1\\
0\\
\end{matrix}  \right)_2
=|V\rangle|+2\rangle
  \label{eq:verK}
\end{align}

\begin{align}
&{\tt L}=\left(
\begin{matrix}
     1\\
     -1\\
     -i\\
     i\\
   \end{matrix}  \right)
\to
\frac{1}{2}\left(
\begin{matrix}
1    \left(
     \begin{matrix}
      1\\
      -1\\
    \end{matrix}  
    \right)
\\
\\
-i   \left(
     \begin{matrix}
      1\\
      -1\\
    \end{matrix}  
    \right)
\\
\end{matrix}  \right)\notag\\
 & =\frac{1}{\sqrt{2}}
  \left(
\begin{matrix}
1\\
-i\\
\end{matrix}  \right)_1
\otimes\frac{1}{\sqrt{2}}\left(
\begin{matrix}
1\\
-1\\
\end{matrix}  \right)_2
  =|L\rangle|v\rangle
  \label{eq:verL}
\end{align}

\bigskip

Thus, in order to handle a complex
coordinatization---states $C$ and $L$---we need a fifth
degree of freedom (circular polarization).

%


\end{document}